\documentclass[aps,prl,preprint, showpacs,groupedaddress]{revtex4}

\usepackage{amsmath}
\usepackage{amssymb}
\usepackage{graphics}
\usepackage[dvips]{graphicx}
\graphicspath{{./},{./pictures/}}
\begin{document}

\title{Quantum Turbulence Decay}

\author{Demosthenes Kivotides}
%\email[]{demos@galcit.caltech.edu}
%\author{}
%\email[]{}
%\thanks{}
\affiliation{Low Temperature Laboratory, Helsinki University of Technology,\\
P.O. Box 2200, FIN-02015 HUT, Finland}

\date{\today}

\begin{abstract}
We develop a computational model of quantum turbulence decay
employing a kinematic prescription for the normal fluid.
We find that after an initial transient, the length of the
vortex tangle $L$ decreases and for large times obeys
the scaling law $L \approx t^{-0.45}$. The average 
magnitude (along the quantized vortices) of the superfluid
and line-vortex velocity are close and differ
significantly from the average magnitude of the normal fluid
velocity.
\end{abstract}

\pacs{67.40.Vs, 47.27.Ak, 47.27.Gs}
%\keywords{}

\maketitle
In quantum turbulence physics \cite{vinen:2002}, a tangle of quantized vortices
interacts via mutual friction forces
with thermal excitations (normal fluid) of the superfluid ground state.
An elementary research program for this multifaceted problem investigates
idealized flows characterized by symmetries like homogeneity
in space and/or time, as well as, isotropy. At first,
phenomenological issues like
scalings of energy spectra \cite{tabeling:1998,kivotides:2002}, 
energy decay \cite{stalp:1999} and 
structure functions should be established. This Letter
contributes to this research program 
by employing a mathematical model of 
decaying quantum turbulence under conditions resembling
the experiment of \cite{stalp:1999} and solving it with numerical
and computational methods.\\ 

Our model consists of a dynamic equation describing the superfluid vortices
and a kinematic prescription for the turbulent velocity field. In particular,
if $\boldsymbol{S}(\xi,t)$ is the three dimensional 
representation of the vortex tangle
then its motion obeys the equation \cite{idowu:2000}:
\begin{eqnarray}
\frac{d\boldsymbol{S}}{dt} = \boldsymbol{V_l}=
&&h \boldsymbol{V_s}+ h_{\times} \boldsymbol{S}^{\prime} \times (\boldsymbol{V_n}-
\boldsymbol{V_s})- \nonumber\\
&&h_{\times \times} \boldsymbol{S}^{\prime} \times (\boldsymbol{S}^{\prime}
\times \boldsymbol{V_n})
\end{eqnarray}
where the superfluid velocity $\boldsymbol{V_s}$ is given by the Biot-Savart integral:
\begin{equation}
\boldsymbol{V_s}(\boldsymbol{x}) = \frac{\kappa}{4 \pi} \int
\frac{(\boldsymbol{S}-\boldsymbol{x}) \times
d\boldsymbol{S}}
{{|\boldsymbol{S}-\boldsymbol{x}|}^3},
\end{equation}
where $t$ is time, $\boldsymbol{x}$ is space, $\kappa$ is the quantum of
circulation, $\boldsymbol{V_n}$ is the velocity
of the normal fluid, $\xi$ is the arclength along the loops, $\boldsymbol{S}^{\prime}=
\frac{d\boldsymbol{S}}{\|d\boldsymbol{s}\|}$ is the unit tangent vector while $h$, 
$h_{\times}$ and $h_{\times \times}$ are constants related to mutual friction physics.\\
At every instant, the normal velocity is decomposed into a mean value
and a fluctuation $\boldsymbol{V_n}= \langle \boldsymbol{V_n} \rangle +
\boldsymbol{u_{n}}$ with the fluctuation $\boldsymbol{u_{n}}$ defined 
by the following function \cite{fung:1998}:
\begin{eqnarray}
\boldsymbol{u_{n}} = &&\sum_{m=1}^{M}
[\boldsymbol{A_m} \times \boldsymbol{\hat{k}_m}
cos(\boldsymbol{k_m} \cdot \boldsymbol{x}+ \omega_m t)+ \nonumber\\
&&\boldsymbol{B_m} \times \boldsymbol{\hat{k}_m}
sin(\boldsymbol{k_m} \cdot \boldsymbol{x}+ \omega_m t)],
\end{eqnarray}
where $M$ is the number of wavemodes the sum of which
constitutes the velocity field. $\boldsymbol{A_m}$, $\boldsymbol{B_m}$ are vectors
with random orientation and magnitude $|\boldsymbol{A_m}|^2 =
|\boldsymbol{B_m}|^2 = (2/3) E_n(k_m) \Delta k_m$ with $E_n(k_m)$ the
normal fluid energy spectrum at wavenumber $k_m$. In addition, 
$\boldsymbol{\hat{k}_m}$ is a unit vector normal to both 
$\boldsymbol{A_m}$ and $\boldsymbol{B_m}$ and 
$\boldsymbol{k_m}=k_m  \boldsymbol{\hat{k}_m}$. The frequencies
$\omega_m = \sqrt{k_{m}^{3} E_n(k_m)}$ correspond to the 
physical notion of ``eddy turnover time''.
\begin{figure*}[t]
\begin{minipage}[t]{0.99\linewidth}
\begin{tabular}[b]{ccc} 
\includegraphics[width=0.33\linewidth]{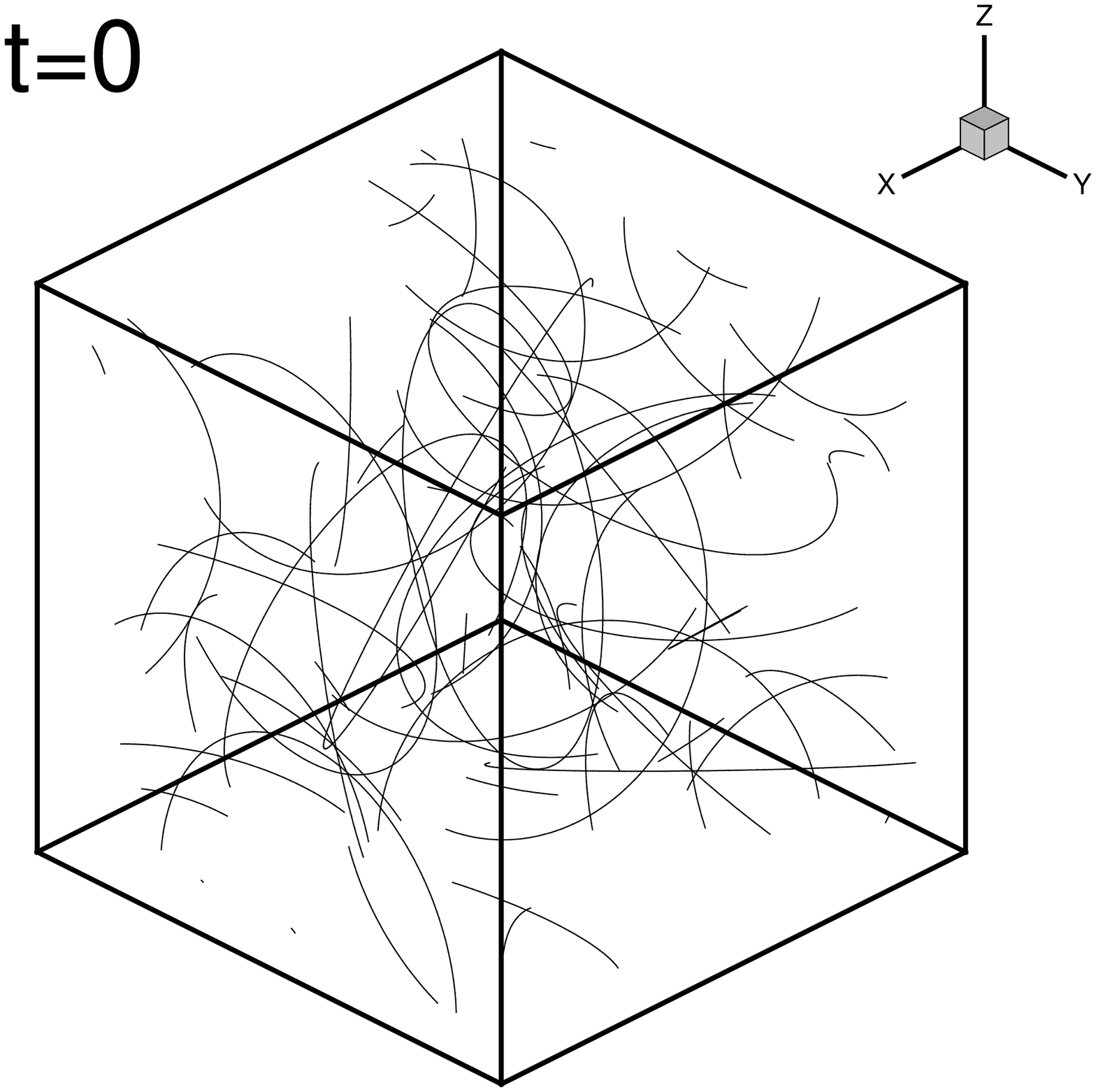} &
\includegraphics[width=0.33\linewidth]{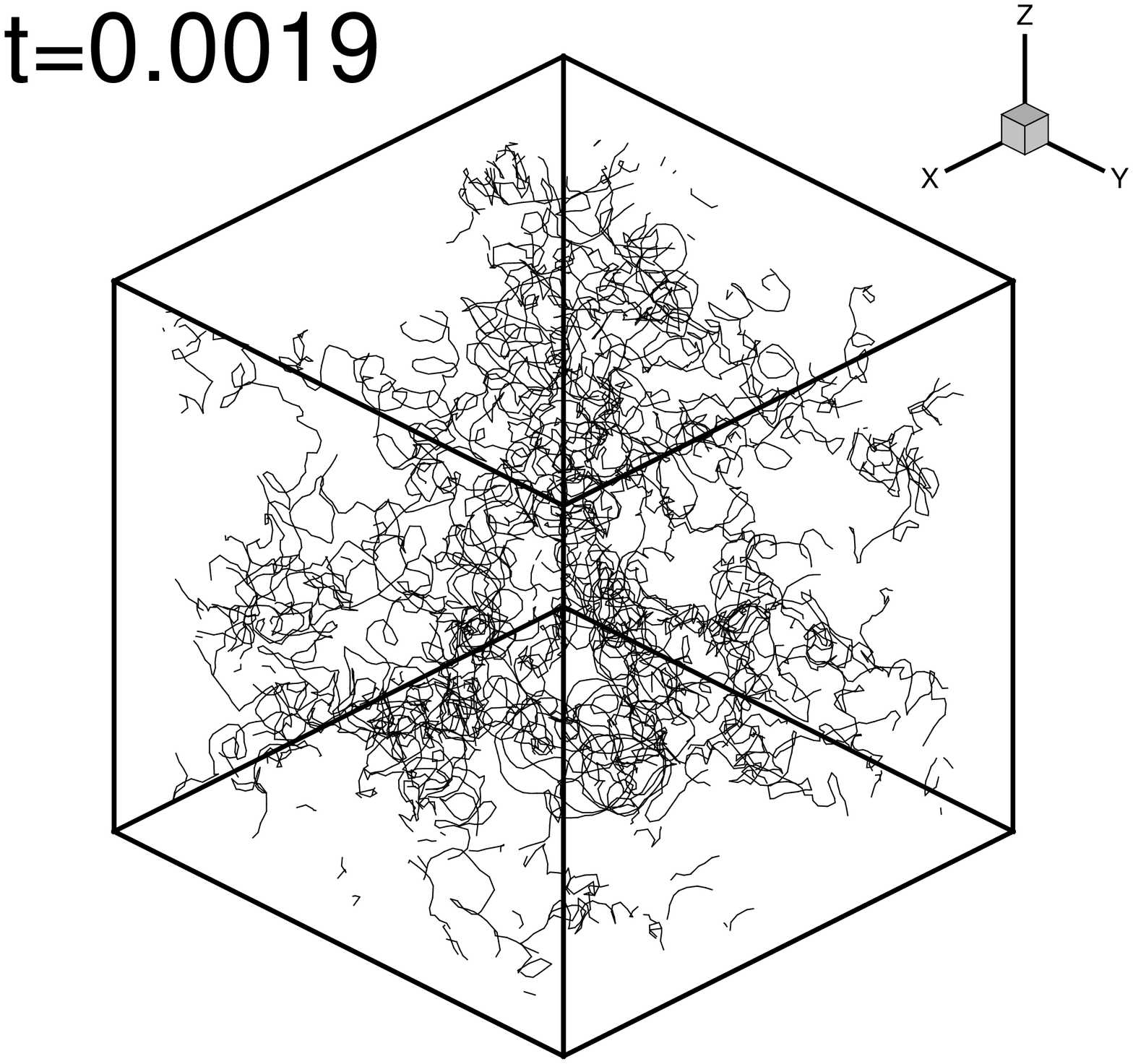} &
\includegraphics[width=0.33\linewidth]{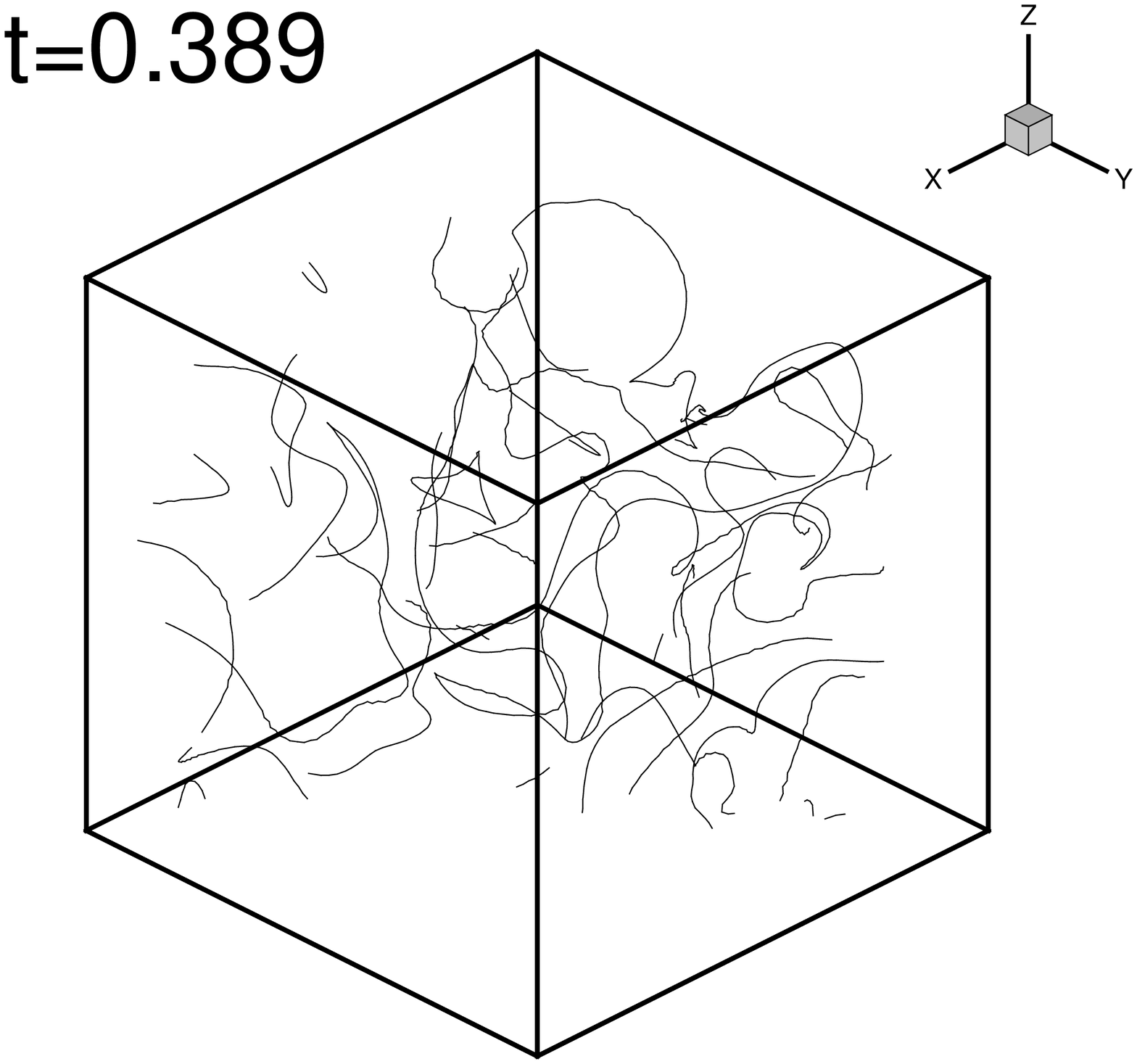}\\
\end{tabular} 
\caption{\label{tangles}
Vortex tangle at initial, maximum length and stoppage times.
To serve clarity, only one eighth
of the computational box is shown.} 
\end{minipage}
\end{figure*}
As required, $\boldsymbol{u_{n}}$
is incompressible by construction.
The energy spectrum $E_n(k_m)$ has two parts: A
high wavenumber part (``inertial range''):
$E_n(k_m)= C_K \epsilon^{\frac{2}{3}} k_{m}^{-\frac{5}{3}}$,
where $C_K=1.5$ is the Kolmogorov constant and
$\epsilon$ is the rate of energy dissipation
and a low wavenumber part (``large eddies''):
$E_n(k_m)= A k_{m}^{2}$,
where $A$ is defined by matching the two spectra at $k_e$ the 
``integral length scale'' wavenumber. According to the
permanence of large eddies hypothesis, $A$ remains constant during the
decay of turbulence.
We observe that in the present 
definition of $\boldsymbol{u_{n}}$
there is no coupling between different modes and so
there can be no energy flux in wavenumber space. This is 
in contrast to what happens in Navier-Stokes turbulence.
In addition, the model is insufficient for capturing
fine normal fluid effects like intermittency.
On the other hand, it reproduces
adequately a number of turbulence phenomenologies 
that relate to even order statistics like
for example the Kolmogorov scaling or 
the Lagrangian flatness factor \cite{malik:1999}. Moreover,
since the model was devised having homogeneous, isotropic turbulence in mind,
it is suitable for the description of the grid normal fluid turbulence
in the \cite{stalp:1999} experiment.\\
We use the symbol $l_e$ for the integral scale of turbulence 
(peak of the spectrum) and the symbol $u_{n}^{\prime}$ for the turbulence
intensity. It is $3 {u_{n}^{\prime}}^2=
\sum_{i=1}^{3} \langle u_{n}^{i} u_{n}^{i} \rangle= 2 E$, 
where $E$ is the kinetic energy
of turbulent fluctuations. Knowledge of $l_e$ and $u_{n}^{\prime}$
at each time step allows the construction of the normal velocity field.
In particular, we can calculate the turbulent Reynolds 
$Re_t= {u_{n}^{\prime} l_e}/{\nu}$ (where $\nu$ is the coefficient
of viscosity), the energy dissipation
rate $\epsilon= {{u_{n}^{\prime}}^3}/{l_e}$
and the Kolmogorov scale $\eta=l_e/ Re_{t}^{{3}/{4}}$.
We employ the decay model of \cite{stalp:1999} in order to
calculate $l_e$ and $u_{n}^{\prime}$ as functions of time.
According to this model, there are two periods of 
turbulence decay: During the first,
$l_e$ is smaller than $l_b$ (the box size). During the second
(which starts at $t_s = \frac{11}{5 {(2\pi)}^{\frac{5}{2}}}
\sqrt{\frac {C_{K}^{3}}{A}} (l_b^{5/2}-l_{e0}^{5/2})$ 
with $l_{e0}$ being the integral scale at $t=0$), 
$l_e$ remains constant and equal to $l_b$.
The change in $l_e$ before its saturation is given by:
$l_e(t)= 2 \pi \bigg{(}{(5/11) 
\sqrt{\frac{A}{C_{K}^{3}}} (t+t_0)}\bigg{)}^{{2}/{5}}$,
with $t_0=\frac{11}{5 {(2\pi)}^{\frac{5}{2}}} 
\sqrt{\frac {C_{K}^{3}}{A}} l_{e0}^{{5}/{2}}$.
For times smaller than the $l_e$ saturation time
the kinetic energy of turbulence $E$ is calculated from:
$E(t)=E_0{(1+\frac{t}{t_0})}^{{-6}/{5}}$,
where $E_0=(9/6){(2\pi)}^{3} A/l_{e0}^{3}$. We have found that
the prefactor $9/6$ is necessary in order not to have a discontinuity in
normal fluid energy at $t_s$. This
condition is not satisfied by the prefactor $11/6$ in formula
$(4)$ of \cite{stalp:1999}.  
For post-saturation times it is:
$E(t)=\frac{27C_{K}^{3}l_{b}^{2}}{2{(2\pi)}^2}
(t+t_0+t_1)^{-2}$,
with $t_1=(4/5){(2\pi)}^{-{5}/{2}} C_{K}^{{3}/{2}}
A^{-{1}/{2}} l_{b}^{{5}/{2}}$.
The constants $t_0$ and $t_1$ define the 
virtual origin time $t_{vo}=-(t_{0}-t_{1})$.
As defined in \cite{skrbek:2000} $t_{vo}$ is
the time when (supposedly) the turbulence has infinite energy 
concentrated on an integral length scale of 
infinite wavenumber. In this interpretation as the energy
decays the energy containing wavenumber moves
towards smaller values. Our initial conditions correspond
to an intermediate turbulence state in the decay process.
We employ periodic boundary conditions for the superfluid
tangle by introducing image vortices. The normal flow
is periodic by construction.\\
The working fluid is $^{4}He-II$
and so the quantum of circulation has the
value $\kappa= 9.97 \cdot 10^{-4} cm^2/s$. The calculation
is done at $T=1.3 K$ (compared with $T=1.5 K$ in
\cite{stalp:1999}) for which the other parameters
of the problem have the values: $\nu= 23.30 \cdot 10^{-4} cm^2/s$,
$h=0.978$, $h_{\times}=4.0937 \cdot 10^{-2}$ and
$h_{\times \times}=2.175 \cdot 10^{-2}$. In addition, we have
$Re_t=5 \cdot 10^3$ and the initial peak of the spectrum is located
at $l_{e0}=0.0161 cm$ which corresponds to $k_{e0}\approx 62 cm^{-1}$.
For comparison, $l_b =0.1 cm$ and $k_{b}=10 cm^{-1}$. We mesh the 
line vortices with discretization length $\Delta x=l_b/84=1.19 \cdot
10^{-3} cm$. We use the same distance to define the smallest
resolvable wavelength in the normal fluid turbulence model (eq. (3)),
$l_{co}=2.381 \cdot
10^{-3} cm$ which corresponds to wavenumber $k_{co}= 420 cm^{-1}$.
Using the equations of the model we can calculate 
the saturation time $t_s=0.2048 \cdot
10^{-2} s$, as well as, the time $t_{\eta r}$ at which $\eta$ 
will become equal to the
smallest resolvable scale $l_{co}$: $t_{\eta r}\approx 0.025 s$.
Beyond this time the line vortices could develop structure at space scales
smaller than the smallest normal turbulence wavelength. At stoppage time
$t_e=0.389 s$, the Kolmogorov scale is equal to 
$5.26 \cdot 10^{-3} cm$ and therefore it is greater than the smallest
resolvable wavelength equal to $2 \cdot \Delta x= 2.38 \cdot 10^{-3} cm$.
The stoppage time Kolmogorov scale corresponds to wavenumber $k_{\eta e}=190 cm^{-1}$.
The time step is chosen in order to ensure that none of the
(resolvable by the numerical grid) Kelvin waves propagates more than
one discretization vortex segment within one calculation step.
The typical time step for this is
$\Delta t \approx 0.3 \cdot 10^{-3} s$. Initially, the tangle consists of
$114$ vortex rings of random orientation and its total length is
$L_0=14.06 cm$. The choice of random initial conditions is justified for two
reasons: (a) there is no experimental information about the actual 
initial tangle configuration which could be employed, (b) since the experimental scalings are
reproducable without an explicit control over the geometry of the initial vortices
it must be the case that the scaling phenomenology does not depend 
on the latter geometry. The second
point agrees with our conception of turbulence as a statistical flow state
that can be achieved from a variety of initial conditions. This is equivalent
to our understanding of turbulence properties as idiosyncratic of 
the differential equations governing the system and not of the initial conditions.
Different initial conditions are driven by the system to generic (reproducable)
turbulence scalings.\\
Other useful quantities one can calculate are the average
values of the velocity magnitudes $|\boldsymbol{V_l}|$, $|\boldsymbol{V_n}|$,
$|\boldsymbol{V_s}|$ and $|cos(\theta)|=
|\boldsymbol{S}^{\prime} \cdot (\boldsymbol{V_n}-\boldsymbol{V_l})|
/|\boldsymbol{V_n}-\boldsymbol{V_l}|$
along the vortex filaments at various times during the system's evolution. 
These averages are taken by sampling the quantities of interest at each discretization node,
and subsequently forming their arithmetic mean. Angle $\theta$
is an important quantity in the physics of the mutual friction
force per unit length: $\boldsymbol{f}= \rho_s \kappa d_{\times \times}
\boldsymbol{S}^{\prime} \times (\boldsymbol{S}^{\prime}
\times (\boldsymbol{V_n}- \boldsymbol{V_l}))-
\rho_s \kappa d_{\times} \boldsymbol{S}^{\prime} \times (\boldsymbol{V_n}-
\boldsymbol{V_l})$. Here $d_{\times}=-2.045 \cdot 10^{-2}$ and
$d_{\times \times}=4.270 \cdot 10^{-2}$
are nondimensional coefficients and $\rho_s=138.6 \cdot 10^{-3} g/cm^{-3}$
is the superfluid density.\\
\begin{figure*}[t]
\begin{minipage}[t]{0.99\linewidth}
\begin{tabular}[b]{cc}
\includegraphics[width=0.49\linewidth]{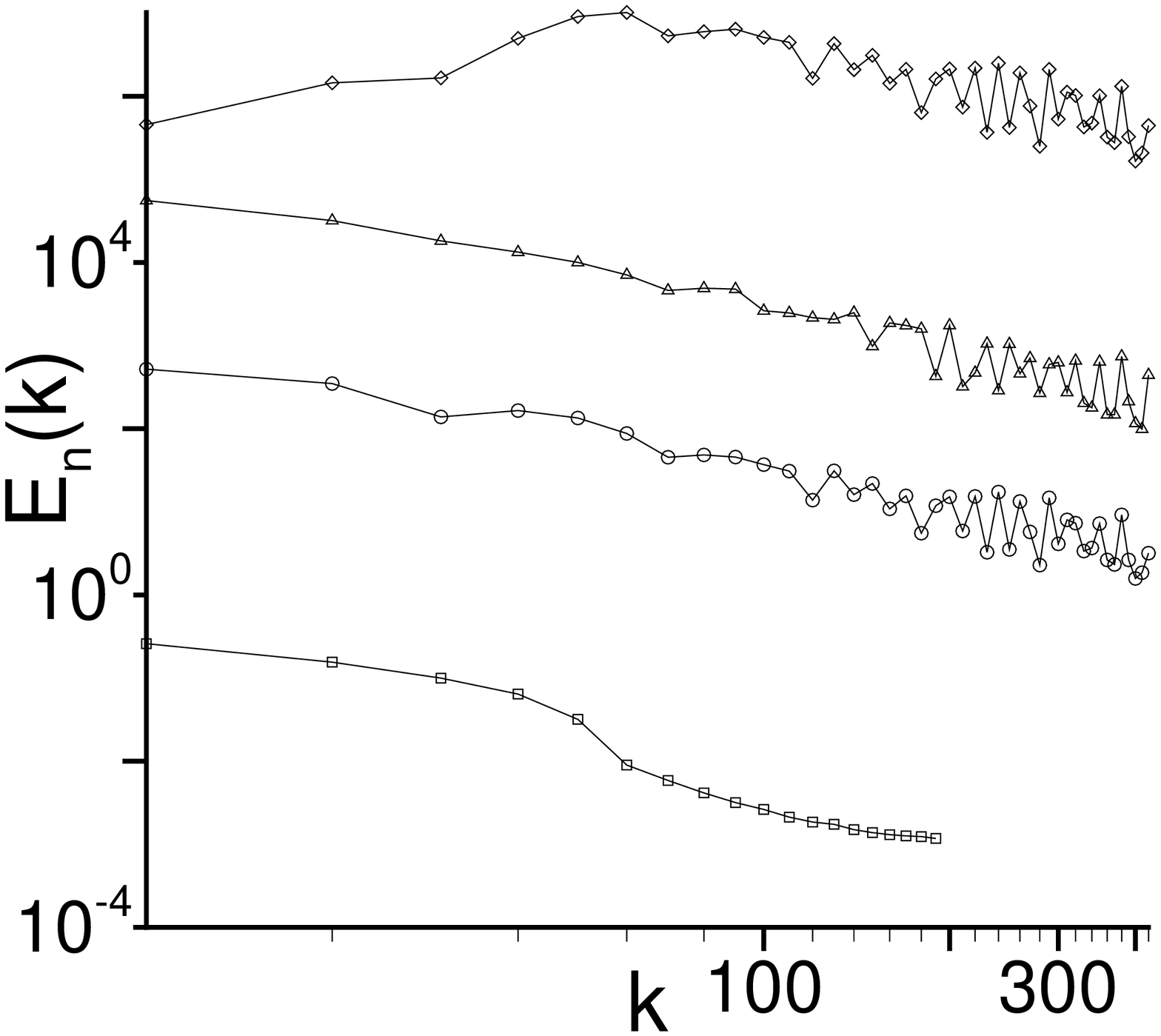} &
\includegraphics[width=0.49\linewidth]{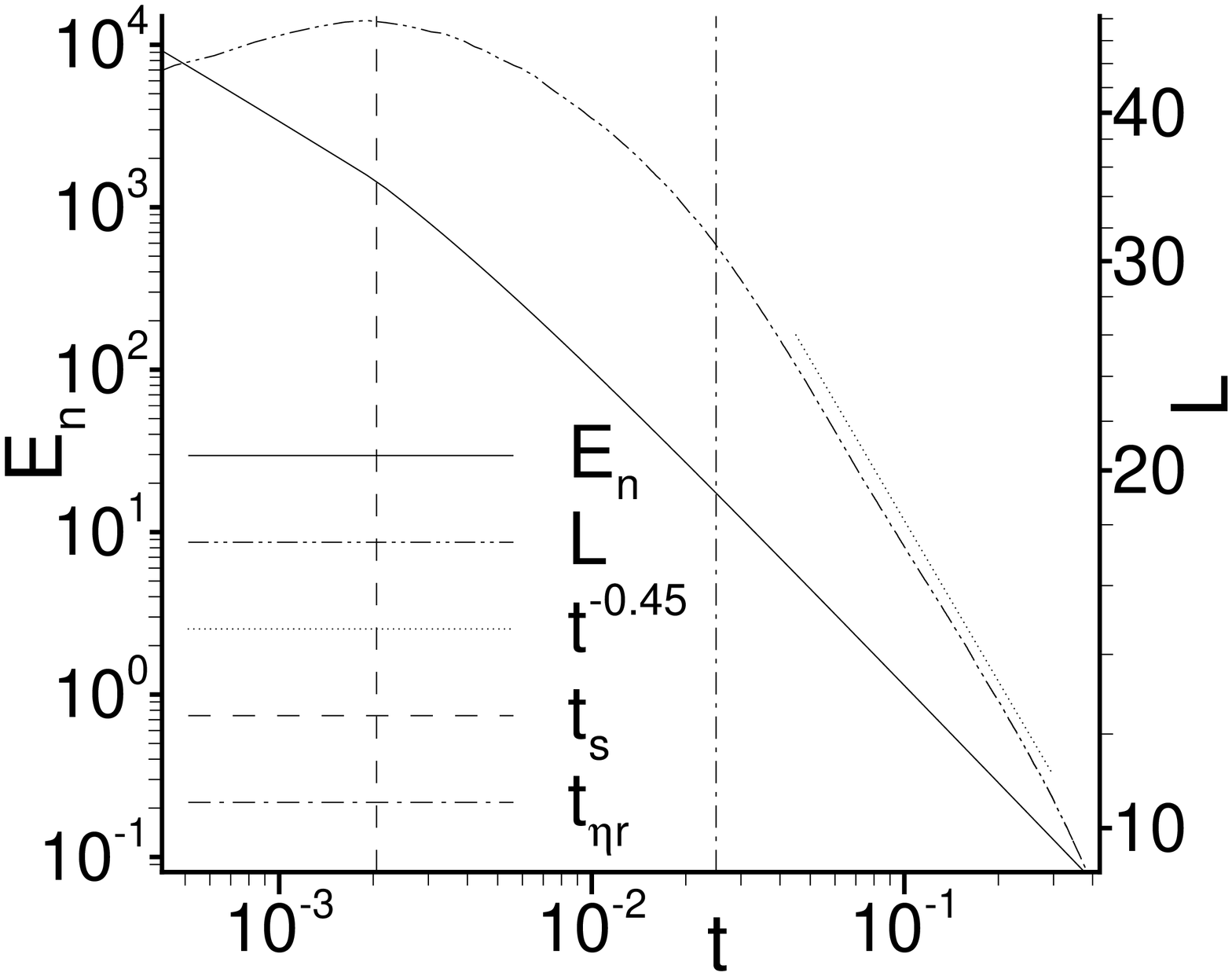} \\
\end{tabular}
\caption{\label{enlen} Left: normal fluid spectra $E_{n}(k)$ at $t=0$, $t=t_s$,
$t=t_{\eta r}$ and $t=t_e$.
Right: the evolution of tangle length $L$ and normal fluid
turbulent energy $E_n$.}
\end{minipage}
\end{figure*}

\begin{figure*}[t]
\begin{minipage}[t]{0.99\linewidth}
\begin{tabular}[b]{cc}
\includegraphics[width=0.49\linewidth]{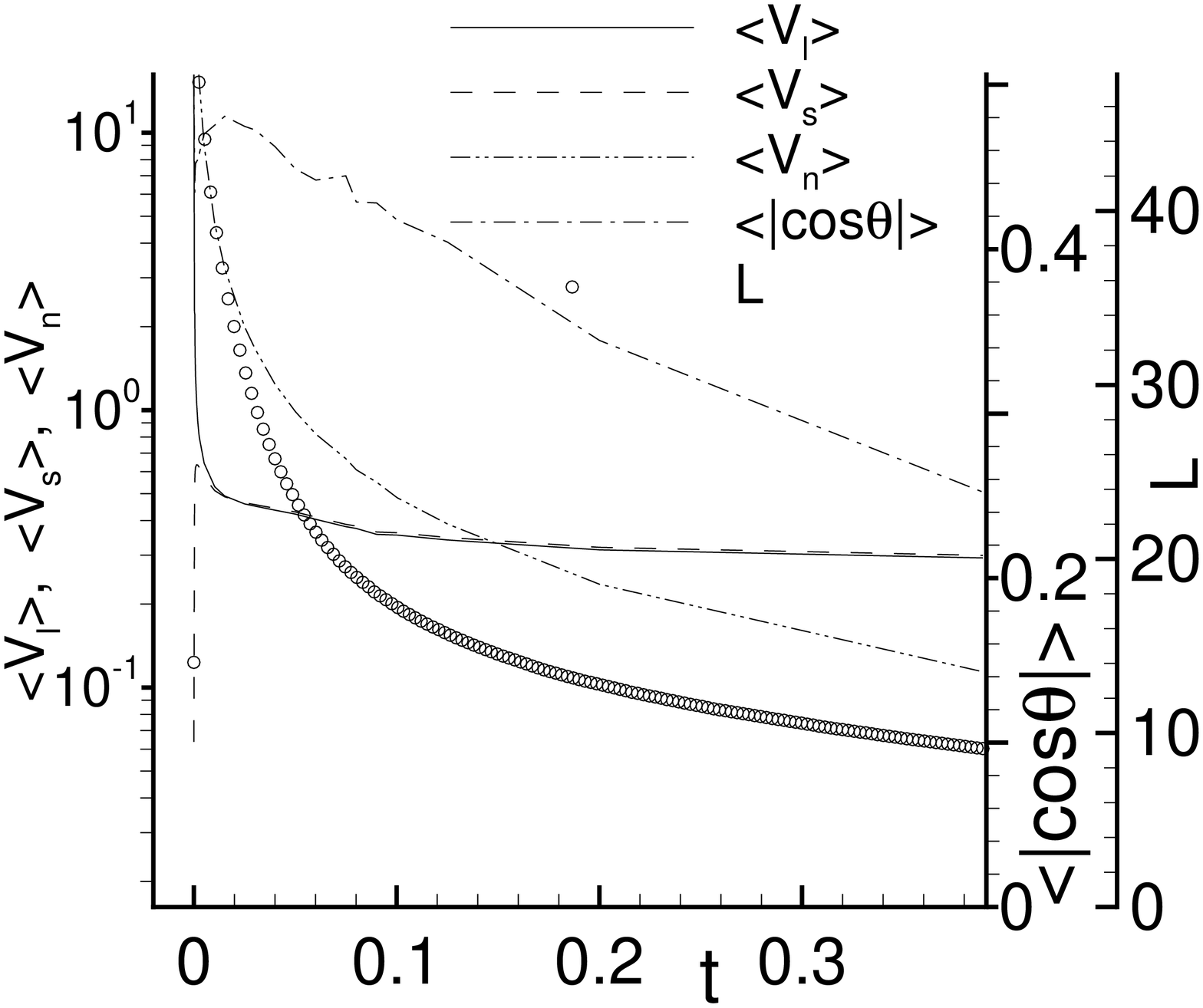} &
\includegraphics[width=0.49\linewidth]{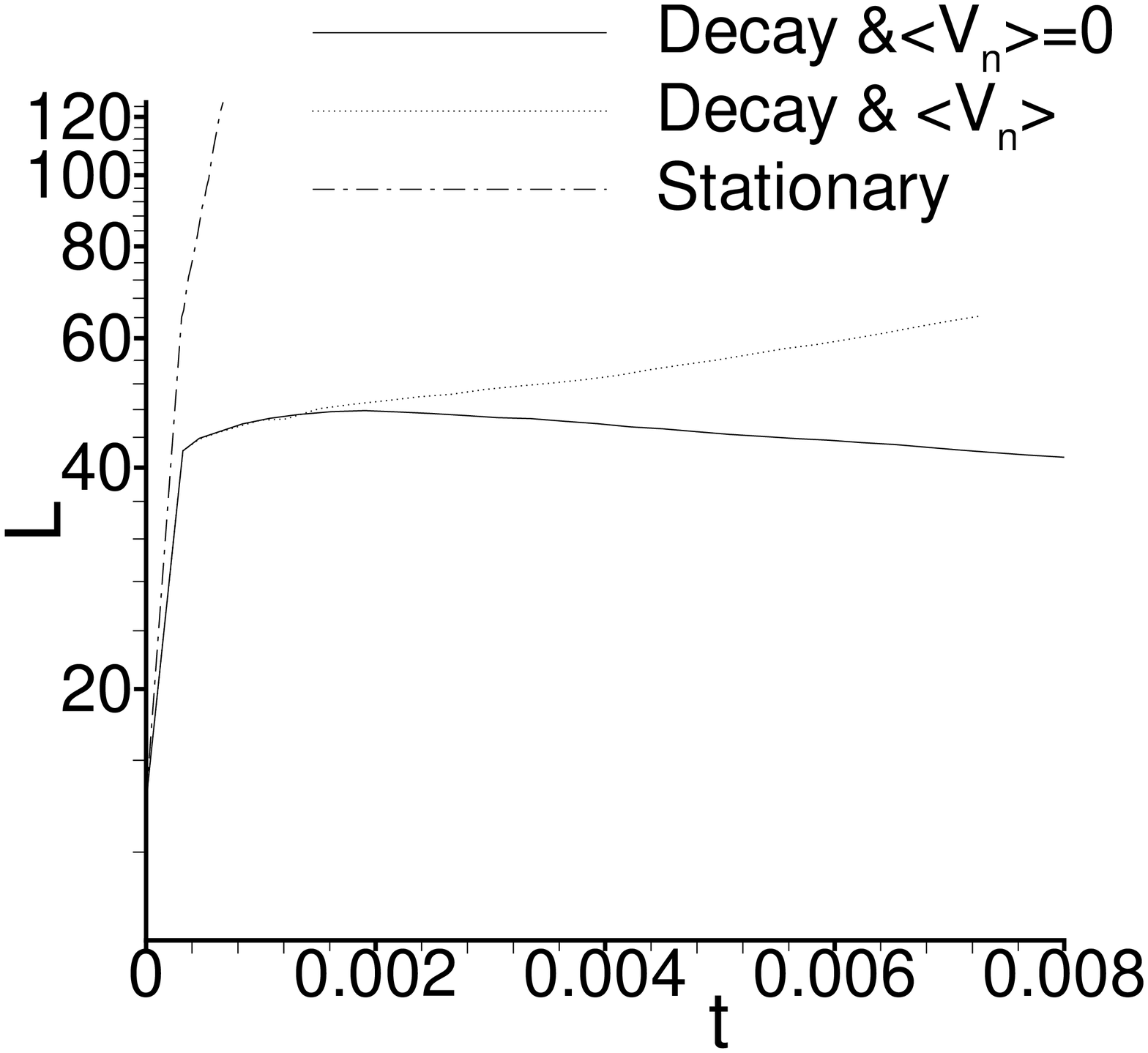} \\
\end{tabular}
\caption{\label{lencomp} Left: average values of $|\boldsymbol{V_l}|$,
$|\boldsymbol{V_n}|$, $|\boldsymbol{V_s}|$ and
$|cos(\theta)|$ along the quantized vortex filaments.
Right: evolution of tangle length $L$ in three calculations with different
normal fluid turbulence characteristics.
}
\end{minipage}
\end{figure*}

The results ($\langle \boldsymbol{V_n} \rangle = \boldsymbol{0}$)
show that in accordance to an instability 
discovered by Cheng $\it{et}$ $\it{al}$ \cite{cheng:1973}
and elaborated mathematically by Glaberson $\it{et}$ $\it{al}$ \cite{glaberson:1974},
the normal flow excites Kelvin waves on the filaments
(Fig.\ref{tangles}).
The length of the tangle reaches a maximum
of $L_{max}=47.83 cm$ at $t=0.0019 s$ and subsequently
decreases. At maximum length the vortex line density is
$\Lambda=L/V\approx 0.5 \cdot 10^5$ compared with 
$\Lambda \approx 2 \cdot 10^5$ for the smallest $\Lambda$ run in
\cite{stalp:1999}. In stating this, we 
made use of the relation $\omega(t)= \kappa \Lambda$ 
in order to deduce $\Lambda$ from 
their superfluid vorticity ($\omega(t)$) data.
Refering again to the smallest $\Lambda$ run in \cite{stalp:1999}, 
we note that $\Lambda$ varies there over $2$ orders of
magnitude while here only by a factor of $5$. 
As required (Fig.\ref{enlen}, right), the decay of turbulent normal 
fluid energy obeys the two previously mentioned temporal scaling laws.
Also demonstrated in the same figure (left) are the two
spatial scaling regimes in the $E_{n}(k)$ spectrum before $t_s$,
as well as, the disappearance of the large eddies scaling
regime for subsequent times. One can ask a two fold
question: (a) why the vortex length
decreases after a rapid transient and (b) why its
observed temporal scaling at large times, $L \approx t^{-0.45}$,
differs from the $L \approx t^{-1.5}$ one of \cite{stalp:1999}?\\
Possible reasons for the latter might be the inadequacy of the employed
turbulence model or the shorter decay time span of $2$ decades in the calculation
compared to $3$ in the experiment. In this milieu, an important question
has to do with the meaning of the reported tangle lengths in both theory
and experiment in the light of the findings of \cite{kivotidesf:2001}
that the superfluid tangle is a fractal. According
to \cite{mandelbrot:1977} (pg $25$), the latter means that, as long
as, the yardstick for length measurements belongs to the scale range within which the 
tangle satisfies a fractal scaling, different yardstick lengths will
give a different length for the tangle. When it comes to calculation,
this implies that better resolved fractal tangles would be measured to posses 
significantly greater lengths when the (different for each resolution)
discretization length is used as yardstick length. Moreover, one could ask 
at first whether the second sound measurement technique introduces (in fractal turbulent
tangles) such a yardstick length depending on the second sound wavelength/frequency.
In other words, whether there exists a certain Kelvin wave 
frequency above which the experimentally employed second sound does 
not see the variations of the line density in a fractal vortex system.
Subsequently, whether this possible uncertainty in the length measurement
affects the scalings observed during the decay of turbulence.
These matters are not clear-cut issues and deserve further 
investigation by expanding (for example) the work of
\cite{samuels:1990} which calculated the motion of one roton shot
toward a single straight line vortex in the realm of interactions between a roton
and fractal vortex lines.\\
It is useful here to notice that in contrast to classical turbulence
a uniform normal velocity field is not compatible
with the assumption of isotropic superfluid turbulence
and it is not dynamically irrelevant.
These are clearly seen in counterflow quantum turbulence calculations
\cite{schwarz:1988} where the superfluid turbulence is
due exclusively to a constant (externally imposed)
velocity field. In these calculations, there is anisotropy
in the direction of the imposed flow.
The above are reminiscent of the role of 
uniform, imposed, magnetic fields
in inducing anisotropies and affecting the velocity spectra
in hydromagnetic turbulence (page $132$ in  \cite{herring:1989}
and page $100$ in \cite{biskamp:2003}). In order to clarify
better this point we have done two more calculations, one with
stationary turbulence and another with turbulence decay but with the addition of
a constant normal velocity field in the $y$ direction. 
We have fixed the velocity magnitude $|\langle \boldsymbol{V_n} \rangle|=
V_{ny}= 7 cm/s$ so that it is comparable to the grid towing velocity
(between $5$ and $200$ $cm/s$) in  \cite{stalp:1999}. Although in this way
$u_{n}^{\prime}/V_{ny} \approx 100$ (at $t=0$) one observes in Fig.\ref{lencomp} (right)
that with the imposed velocity field the vortex length increases at times for which
(in the case of purely decaying turbulence) it decreases. Therefore,
the temporal decay law for the line-vortex length could be affected by 
small bulk normal fluid velocities. We implicitly assume here that the mean
velocity profile is stable; this is also the case in \cite{schwarz:1988}.\\
To analyze the first leg of the previously posed question, 
we note that according to the analysis of 
the \cite{cheng:1973,glaberson:1974} instability
(see also the discussion in \cite{kivotides:1999}) the normal fluid
velocity fluctuations transfer energy to a Kelvin wave of a particular
wavelength (and therefore increase its amplitude)
only when their component along the direction of motion of the wave
is both greater in magnitude than the group velocity of the wave
and parallel (of the same sign) to it. Any normal fluid
velocity antiparallel to a vortex wave reduces its amplitude. 
Therefore since initially the turbulence
intensity is approximately $100$ times the group velocity of the
fastest (resolvable) Kelvin wave and there are
no Kelvin waves present, there is unhindered Kelvin wave
excitation. This explains the initial rapid increase of vortex length.
Subsequently, in conjunction with the decay of turbulence intensity,
at places with wave group velocity (a) larger than
the local normal fluid velocity magnitude or (b) antiparallel to the
normal fluid velocity direction, the wave amplitudes are damped.
The turbulence  decay factor is necessary 
since as Fig.\ref{lencomp} (right) shows, in stationary
turbulence the length keeps increasing rapidly.\\ 
The results for the average values of $|\boldsymbol{V_l}|$,
$|\boldsymbol{V_n}|$, $|\boldsymbol{V_s}|$ and
$|cos(\theta)|$ are presented in Fig.\ref{lencomp} (left).
They show that the initial length transient corresponds
to rapid $\langle |\boldsymbol{V_l}| \rangle$ and 
$\langle |\boldsymbol{V_s}| \rangle$
transients. The latter increases towards a maximum
coinciding in time with the length maximum while the former
decreases approaching $\langle |\boldsymbol{V_s}| \rangle$. Notice
that when the length starts decreasing there is order
of magnitude difference between $\langle |\boldsymbol{V_n}| \rangle$ and
$\langle |\boldsymbol{V_s}| \rangle$. This is particularly true for 
the average of $|\boldsymbol{V_s}|$ taken over the whole
volume of the fluid since this was found (using a $84^3$ grid)
to be an order of magnitude smaller than the average over the 
line vortices. Another observation is that for $\langle |\boldsymbol{V_n}| \rangle$ 
smaller than $\langle |\boldsymbol{V_s}| \rangle$ (which happens at very small 
normal turbulence energies) $\boldsymbol{S}^{\prime}$ tends to become normal to 
$\boldsymbol{V_n}-\boldsymbol{V_l}$. The results bring forward
quantum turbulence physics 
that differ significantly from those proposed in \cite{stalp:1999}
where the assumption was made that the superfluid and normal
fluid velocities are identical. However, one must also bear in mind
that the employed turbulence model does not have the quality of
fully dynamical Navier-Stokes calculations. The latter 
kind of computations would eventually be required in 
order to verify the present results.\\

Overall, it is not possible to argue for the
quality of the established mathematical models
in quantum turbulence theory, as long as, the 
available experimental data fail to address directly
the main variables (fluid velocities, vortex tangle
geometry) in these models. This is a major
obstacle for progress in quantum turbulence research.
The present work suggests that progress depends on
the development of new, more potent
experimental methods and their combination with fully dynamic
mathematical calculations. The latter could identify generic
and essential phenomenological trends that could be encoded in
statistical mechanical equations.
\begin{acknowledgments}
This research was supported by the Commission of the European Union
under Contract \# HPRI-CT-1999-00050. I thank Matti Krusius and Carlo Barenghi  
for discussions, as well as, Caltech for computing time.
\end{acknowledgments}
\bibliography{decay}
\end{document}